\documentclass[12pt,a4paper]{article}
\usepackage{amssymb}
\usepackage[T2A]{fontenc}
\usepackage[cp866]{inputenc}
\usepackage[russian,english]{babel}
\usepackage{graphicx}
\usepackage[dvips]{psfrag}
\usepackage[unicode]{hyperref}
\begin{document}
\begin{center}
{\Large What is the quantity dual to the Lagrangian density?}\\[3mm]
{B.~P.~Kosyakov}\\[3mm]
{{\small Russian Federal Nuclear Center--VNIIEF, Sarov, 607188 Nizhni{\u\i} Novgorod Region, Russia;\\
Moscow Institute of Physics {\&} Technology, Dolgoprudni{\u\i}, 141700 Moscow Region, Russia}} 
\end{center}
\begin{abstract}
\noindent
{We address the lately discovered Lagrangian density ${\cal L}$ of nonlinear electrodynamics 
preserving both conformal invariance and  electric-magnetic duality to show that the quantity dual 
to ${\cal L}$ is ${\cal K}=\frac12\sqrt{\Theta_{\mu\nu}\Theta^{\mu\nu}}$, where $\Theta_{\mu\nu}$ 
is the stress-energy tensor built out of this ${\cal L}$.
We point out that ${\cal L}$ and ${\cal K}$ make up a pair of canonically conjugate variables
which can be regarded as local coordinates of a two-dimensional symplectic manifold.}
\end{abstract}
In modern theoretical physics we feel the need for what amounts to a variable conjugate to the 
Lagrangian ${\cal L}$, or else a quantity dual to ${\cal L}$ in some sense.
To illustrate, we refer to quantum gravity where the functional integration is easiest to treat in 
the Lagrangian formalism because the terms contributing to ${\cal L}$ are associated with geometric 
invariants, while the Feynman integral is normally defined in a phase space framework.
The present letter makes a step toward the development of the desired concept. 
Our main concern here is with nonlinear generalizations of Maxwell's electrodynamics which enjoy 
the properties of conformal invariance and electric-magnetic duality.

It may appear that the variable conjugate to ${\cal L}$ should be attributed to the metric 
stress-energy tensor  $T^{\mu\nu}$ whose construction
\footnote{The derivation of most algebraic relations given below without proof can be found, e.~g., 
in \cite{Kosyakov}.} 
\begin{equation}
T^{\mu\nu}=2\,\frac{\partial{\cal L}}{\partial g_{{\mu\nu}}}-g^{{\mu\nu}}\,{\cal L}\,
\label{stress-energy_tensor}
\end{equation}
vaguely resembles the Legendre transform of ${\cal L}$.
However, ${\cal L}$ and $T^{\mu\nu}$ drastically differ in their transformation properties.
We therefore will not try to offer universal definitions, and restrict our consideration to the 
theories patterned after electrodynamics.

Nonlinear versions of electrodynamics preserving both electric-magnetic duality and conformal 
invariance were shown \cite{Bandos, K-20} to be encoded in the one-parameter family of Lagrangian
densities
\footnote{Heaviside units are adopted throughout.}    
\begin{equation}
{\cal L}({\cal S},{\cal P};{\alpha})=-\frac12\left({\cal S}\,\cosh \alpha-
\sqrt{{\cal S}^2+{\cal P}^2}\,\sinh \alpha\right),
\label
{final-solution}
\end{equation} 
where ${\cal S}=\frac12\,F_{\mu\nu}F^{\mu\nu}$ and ${\cal P}=\frac12\,F_{\mu\nu}{}^\ast\! F^{\mu\nu}$ 
are the electromagnetic field invariants, and $\alpha$ is a variable running from $0$ to $\infty$ 
to parametrize the family.
$\alpha=0$ refers to the free Maxwell electrodynamics governed by the Larmor Lagrangian 
${\cal L}_{\rm L}=-\frac12\,{\cal S}$.

The stress-energy tensor of electromagnetic field is given by
\begin{equation}
\Theta_{\lambda\mu}=-F_{\lambda\sigma}E^{\sigma}_{~\mu}-g_{\lambda\mu}\,{\cal L}\,,
\label{Max-stress-energy}
\end{equation}
where $E_{\mu\nu}$ stands for the field excitation,
\begin{equation}
E_{\mu\nu}=\frac{\partial{\cal L}}{\partial F^{\mu\nu}}\,.
\label
{E-df}
\end{equation}                          
In Maxwell's electrodynamics, $E_{\mu\nu}=-F_{\mu\nu}$, and (\ref{Max-stress-energy}) takes the 
familiar form:
\begin{equation}
\Theta_{\lambda\mu}=F_{\lambda\sigma}F^{\sigma}_{~\mu}+\frac12\,g_{\lambda\mu}\,{\cal S}\,.
\label{Maxw-stress-energy}
\end{equation}
This expression for $\Theta_{\lambda\mu}$ is subject to the relation:
\begin{equation}
4\,\Theta_{\mu\sigma}\Theta^{\nu\sigma}=\left({\cal S}^2+{\cal P}^2\right)\delta_\mu^{~\nu}\,. 
\label
{Max-stress-energy-rel}
\end{equation}

Let us define the quantity
\begin{equation}
{\cal K}=\frac12\sqrt{\Theta_{\mu\nu}\Theta^{\mu\nu}}\,,
\label{K-def}
\end{equation}
and take a closer look at it as applied to the Lagrangian (\ref{final-solution}).  
Combining (\ref{E-df}) and (\ref{final-solution}), 
\begin{equation}
E^{\mu\nu}=\left(-\cosh \alpha+\frac{{\cal S}\,\sinh \alpha}{\sqrt{{\cal S}^2+{\cal P}^2}}\right)F^{\mu\nu}
+\frac{{\cal P}\,\sinh \alpha}{\sqrt{{\cal S}^2+{\cal P}^2}}\, {}^\ast\!F^{\mu\nu}\,,
\label
{ast-G-df}
\end{equation}                          
substituting this into (\ref{Max-stress-energy}), and having regard to the fact that 
$F_{\mu\sigma}{}^\ast\! F^{\sigma\nu}=-\frac12\,{\cal P}\delta_\mu^{~\nu}$, gives
\begin{equation}
\Theta_{\lambda\mu}=\left(\cosh \alpha-\frac{{\cal S}\,\sinh \alpha}{\sqrt{{\cal S}^2+{\cal P}^2}}\right)
\left(F_{\lambda\sigma}F^{\sigma}_{~\mu}+\frac12\,g_{\lambda\mu}\,{\cal S}\right). 
\label
{Theta-alpha}
\end{equation} 
By (\ref{Max-stress-energy-rel}), 
\begin{equation}
\Theta_{\mu\nu}\Theta^{\mu\nu}=\left(\sqrt{{\cal S}^2+{\cal P}^2}\,\cosh \alpha-
{\cal S}\,\sinh \alpha\right)^2.
\label
{Theta-sqr}
\end{equation} 
                    
Now (\ref{final-solution}) and (\ref{Theta-sqr}) can be rewritten as
\begin{equation}
{\cal L}({\alpha})={\cal L}(0)\,\cosh \alpha+{\cal K}(0)\,\sinh \alpha\,,
\label
{L-alpha-through-L-K}
\end{equation} 
\begin{equation}
{\cal K}(\alpha)={\cal K}(0)\,\cosh \alpha+{\cal L}(0)\,\sinh \alpha\,.
\label
{Theta-sqrt}
\end{equation} 

These equations have much in common with a Lorentz boost of fictitious `time' and `space' 
coordinates ${\cal L}$ and ${\cal K}$ 
\footnote{Here, $\cosh\alpha$ plays the role of the Lorentz factor $\gamma=\left(1-V^2\right)^{-1/2}$, in
particular $\cosh\alpha\to\infty$ as $\alpha\to\infty$ much like the transition to a frame of
reference moving at the speed of light, $\gamma\to\infty$ as $|V|\to 1$.}. 
The dissimilarity however is that (\ref{L-alpha-through-L-K})--(\ref{Theta-sqrt}) do not form a 
transformation group unless the parameter $\alpha$ is continued to negative values.
Allowing for negative $\alpha$, one leaves room for an unbounded range of ${\cal L}$, which in turn
renders the Wick rotation of no use for the standard path integral technique.
Meanwhile, in a higher level theory where ${\cal L}$ and ${\cal K}$ become coordinates of a manifold 
which is the arena of a new quantum dynamics, the extension of $\alpha$ to the whole real axis 
${\mathbb R}$ may seem well founded.
This fall of the status of ${\cal L}$ and ${\cal K}$ is somewhat similar to the situation that the
operator ${\hat{\bf r}}$ in the first quantization picture turns to a mere $c$-number coordinate 
${\bf r}$ in the second quantization picture.
We assume that $\alpha\in {\mathbb R}$, and thereby promote the algebraic relations 
(\ref{L-alpha-through-L-K})--(\ref{Theta-sqrt}) to the O$(1,1)$ group transformations. 

If it is granted that ${\cal L}$ and ${\cal K}$ are rectilinear local coordinates of a 
two-dimensional Lorenzian manifold, the question arises: Does the concept of light cone as 
an unpenetrable barrier still stand? 
The invariant quantity is ${\cal K}^2-{\cal L}^2={\cal P}^2$, which suggests that ${\cal P}^2=0$ 
could be likened to the light cone.
Indeed, the Gaillard--Zumino criterion \cite{GaillardZumino} 
\begin{equation}
{}^\ast\! E_{\mu\nu}\,E^{\mu\nu}={}^\ast\!F_{\mu\nu}\,F^{\mu\nu}\,
\label
{duality-cond}
\end{equation}                                    
for invariance under the general electric-magnetic duality rotations
\begin{equation}
E'_{\mu\nu}=E_{\mu\nu}\cos\theta+{}^\ast\! F_{\mu\nu}\sin\theta,
\quad
{}^\ast\!F'_{\mu\nu}={}^\ast\! F_{\mu\nu}\cos\theta-E_{\mu\nu}\sin\theta\,
\label
{duality-transf}
\end{equation} 
is singular at ${\cal P}=0$.
We therefore discriminate between `timelike' regions, ${\cal P}^2>0$, specified by real-valued 
${\cal L}$'s, and `spacelike' regions, ${\cal P}^2<0$, in which complex-valued ${\cal L}$'s  
are acceptable.
To specify the line of demarcation between these regions, we note that 
${\cal P}^2=4\,{\rm det}\left(F_{\lambda\mu}\right)$, which implies that the action 
\begin{equation}
\int d^4 x\,\sqrt{{\rm det}\left(F_{\lambda\mu}\right)}
\label
{int-sqrt-det-F}
\end{equation} 
does not depend on the metric of spacetimes,  and hence gives rise to a topological field theory.
It transpires that going from the `timelike' region to the `light cone' is destructive for either 
conformal symmetry or electric-magnetic duality (\ref{duality-transf}) even at the classical 
level~\footnote{Where the divergency of $\cosh\alpha$ as $\alpha\to\infty$, displayed in 
(\ref{L-alpha-through-L-K})--(\ref{Theta-sqrt}), signals about this danger.}, 
not to mention the affair at the quantum level where a pertinent anomaly may occur.

Since ${\cal L}$ may be arbitrarily small, we are entitled to consider the infinitesimal quantities 
$d{\cal L}$ and $d{\cal K}$.
Clearly their transformation law is identical to that given by
(\ref{L-alpha-through-L-K})--(\ref{Theta-sqrt}),
\begin{equation}
d{\cal L}'=d{\cal L}\,\cosh\alpha+d{\cal K}\,\sinh\alpha\,,
\label
{dL-alpha-through-L-K}
\end{equation} 
\begin{equation}
d{\cal K}'=d{\cal K}\,\cosh \alpha+d{\cal L}\,\sinh\alpha\,.
\label
{dTheta-sqrt}
\end{equation} 
Note that O$(1,1)\sim{\rm Sp(1)}$. 
Hence, (\ref{dL-alpha-through-L-K})--(\ref{dTheta-sqrt}) can be regarded as canonical 
transformations leaving invariant the 2-form $d\Omega=d{\cal L}\wedge d{\cal K}$.
It has long been known \cite{Zeeman} that a proper definition of the topology inherent in 
four-dimensional Minkowski space is a challenge.
In contrast, the natural topology of two-dimensional Minkowski space is easy devised with the aid
of the two-dimensional phase space volume $d\sigma=d\Omega$ invariant under ${\rm Sp(1)}$ 
transformations (\ref{dL-alpha-through-L-K})--(\ref{dTheta-sqrt}).  

Supersymmetric generalizations of ${\cal L}$ defined in (\ref{final-solution}) were recently 
proposed in \cite{Kuzenko, BandosSUSY}.
It would be interesting to clarify if the results of the present letter are at least partially 
reproducible in 
supersymmetric versions of conformally invariant and electric-magnetic duality invariant 
electrodynamics.    

I thank Paul Townsend for useful remarks.

\end{document}